\documentclass[pra,showpacs,nobibnotes,floatfix,superscriptaddress]{revtex4}

\usepackage{latexsym,amsmath,amssymb,amsfonts,mathbbol,graphicx,color}
\usepackage{dcolumn}
\usepackage{bm}
\usepackage{graphicx}
\usepackage{epstopdf}

\begin{document}

\title{Imperfect Construction of Microclusters}

\author{E. Schneider, K. Zhou, G. Gilbert, Y.S. Weinstein}
\affiliation{Quantum Information Science Group, {\sc Mitre},
200 Forrestal Rd. Princeton, NJ 08540, USA}

\begin{abstract}
Microclusters are the basic building blocks used to construct cluster states capable of supporting fault-tolerant quantum computation. In this paper, we explore the consequences of errors on microcluster construction using two error models. To quantify the effect of the errors we calculate the fidelity of the constructed microclusters and the fidelity with which two such microclusters can be fused together. Such simulations are vital for gauging the capability of an experimental system to achieve fault tolerance.
\end{abstract}


\maketitle

\section{Introduction}

Cluster state quantum computation is a quantum computing paradigm that can be implemented by applying single-qubit measurements only to a `cluster state' \cite{BR1,BR2,BR3}. The cluster state is a highly entangled state in which qubits initially in the state $|+\rangle = \frac{1}{\sqrt{2}}|0\rangle+|1\rangle$ are entangled via controlled-PHASE gates. Measurements of the qubits along axes in the \emph{x-y} plane implement the desired algorithm. A cluster state with qubits arranged as a two-dimensional array with nearest neighbor entanglement suffices for universal quantum computation. 

A possible experimental venue for cluster states is photonics. Nielsen noted \cite{N} that by using non-deterministic elements of linear optics quantum computation for photonic cluster construction cluster state quantum computation may be more efficient than optical circuit-based quantum computation. Browne and Rudolph \cite{BR} refined this idea replacing Nielsen's construction method with simpler, also probabilistic, `fusion' operations. A number of additional methods for constructing clusters have been suggested \cite{DR,CCWD} and small photonic cluster states have been experimentally constructed and reported on \cite{Zeil,K,Pan}. 

Photonic cluster state quantum computing, as true for all quantum computing schemes, is subject to errors. Thus, fault tolerant schemes, allowing for successful computation despite errors in the basic components of the computer, must be devised. One such scheme was proposed by Dawson, et. al. \cite{DHN}. This scheme utilizes certain error correction techniques borrowed from circuit-based quantum computation along with some novel ideas on syndrome measurement. The cluster states themselves are constructed using fusion operations. 

Browne and Rudolph \cite{BR} utilized two types of fusion operations to build arbitrary cluster state. Type 1 fusion operations were originally introduced to fuse together cluster chains (clusters in which the qubits are arranged in a 1-dimensional array) and Type 2 fused chains into an `L'-shape. Later it was shown that Type 1 fusion alone allows for the construction of two-dimensional clusters \cite{us} and thus we explore only this type of fusion operation. Fusion is implemented using a polarizing beam-splitter (PBS), a waveplate, and a photon detector, as shown in Fig.~1. Two photons, an edge photon of a cluster chain length $m$ and an edge photon of a chain of length $n$ enter the PBS at adjacent sides. Each photon will, with probability .5 reflect, or, with probability .5, transmit. Thus, with a probability of .5, one and only one photon will emerge from the PBS towards the photon detector. That photon will be rotated by $45^\circ$ and detected fusing the two chains into a single chain of length $m+n-1$. However, with probability .5 zero or two photons will be detected. In this case the two photons are lost, fusion has failed and the input cluster chain lengths are reduced to $m-1$ and $n-1$. 

\begin{figure}
\begin{center}
\includegraphics[height=4cm]{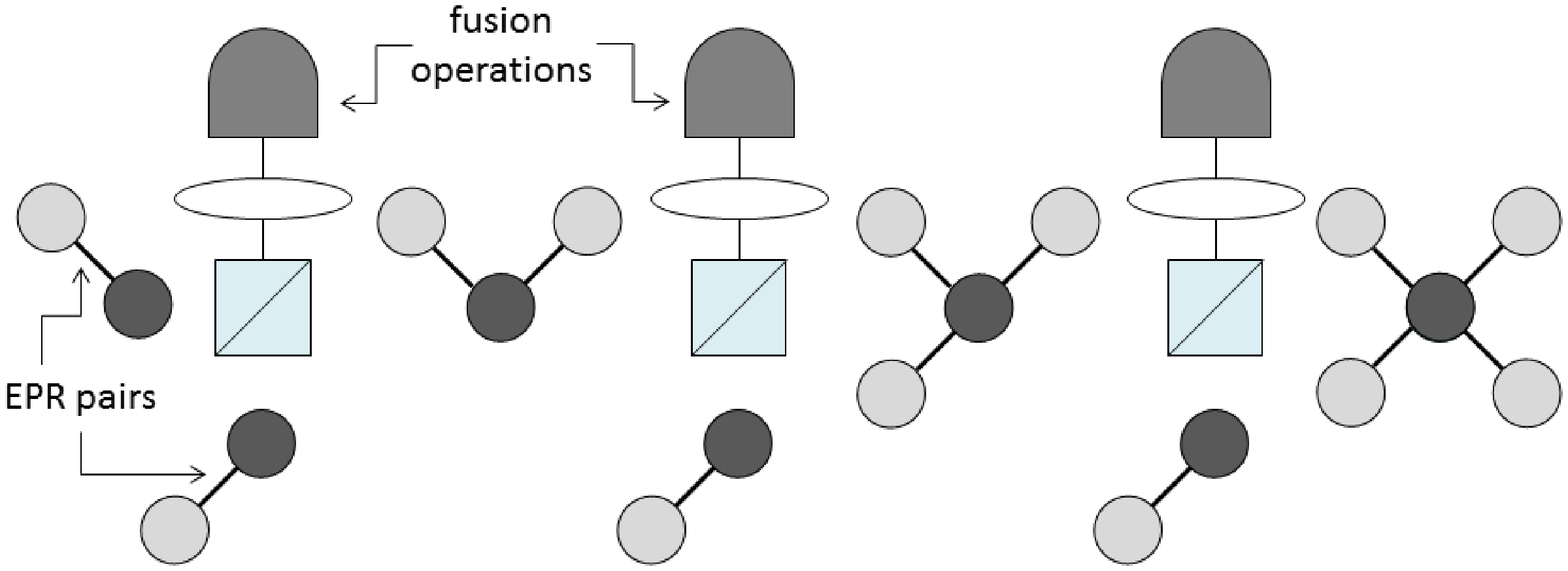}
\includegraphics[height=4cm]{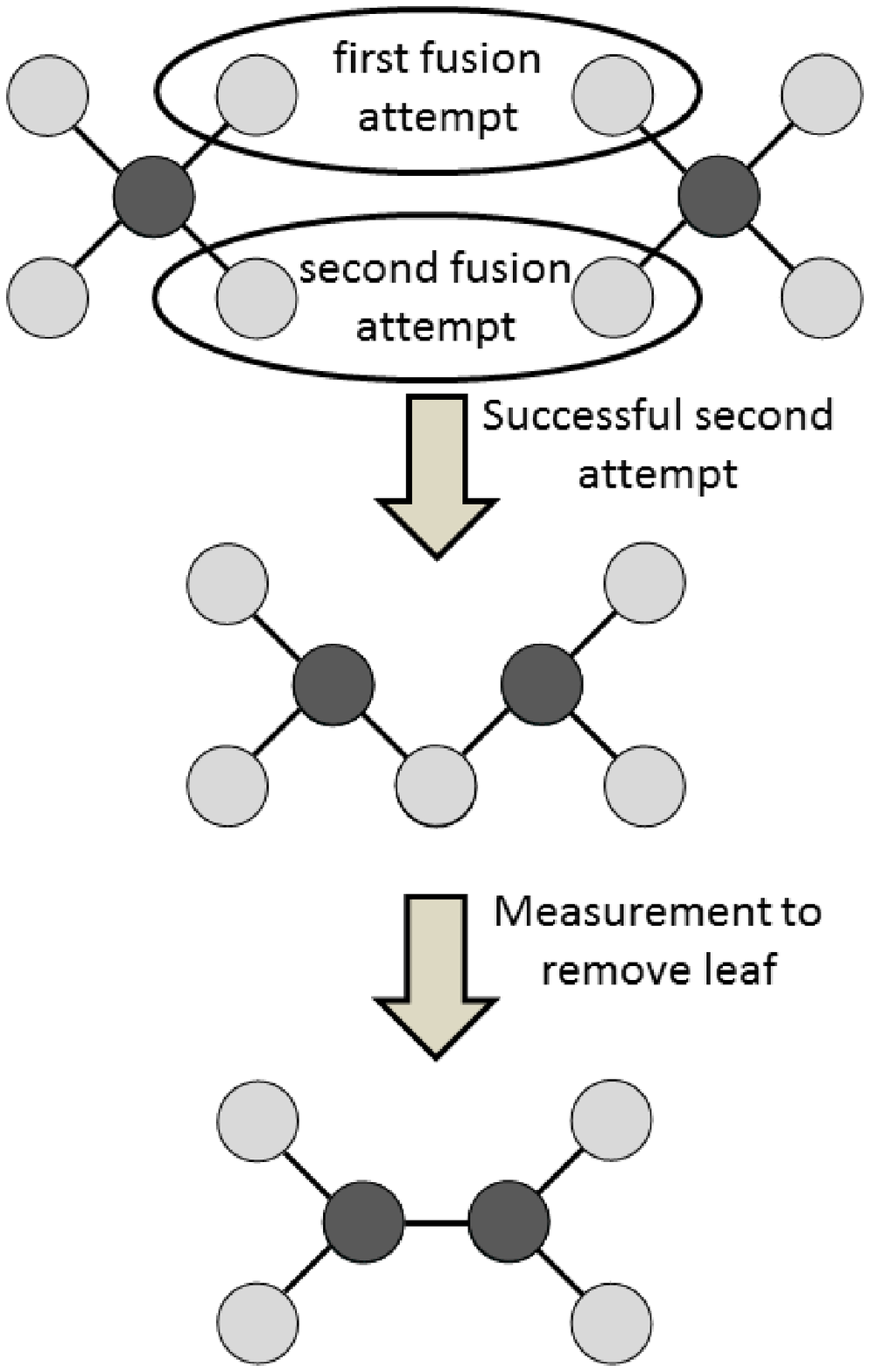}
\caption{\label{micro}
Left: microcluster construction via fusion operations. Two EPR pairs are fused together to form a two-leaf microcluster (leaf nodes light, root nodes dark). The root node is then fused with additional EPR pairs to grow more leaves.
Right: two microclusters are fused via their leaf nodes. If fusion fails the leaf nodes are lost but fusion can be attempted with remaining leaves. Upon success a $y$-axis measurement of the connecting leaf node removes it while joining the two root nodes. }
\end{center}
\end{figure}

To maximize the probability of successful fusion between two information carrying photons Dawson, et. al. \cite{DHN} introduced `microclusters.' Each microcluster is composed of an information carrying root node from which sprout several leaf nodes. Microclusters are fused together without risk of losing the information stored in the root nodes by fusing pairs of leaf nodes. The joining of two micrclusters via a fusion operation are depicted in the Figure \ref{micro}.

To construct a microcluster we start with an unlimited number of polarization entagled pairs of photons, also called Einstein-Podolsky-Rosen (EPR) pairs, assumed to be perfectly generated via a process such as spontaneous parametric downconversion. Two such pairs are fused together generating a three photon cluster. The middle one of the three photons, the one that survived the fusion operation, is the root node. The root node is then fused with a photon from another EPR pair producing a microcluster with three leaves. Additional successful fusion operations will add more leaves onto the root node as shown in Fig.~\ref{micro}. Of course, if any one of the fusion operations is unsuccessful the entire microcluster is destroyed. Nevertheless, since an arbitrarily large number of attempts can be done in parallel, a $k$-leaf microcluster can, with probability arbitrarily close to one, be constructed in time of order ${\rm{log}}k$ using of order $k^2$ EPR pairs. 

As explained above, success of a fusion operation is signified by one photon being detected. However, this does not mean that the fusion operation is done perfectly. In reality the mode-matching and hardware are not perfect. Thus, the success signal from a fusion operation, while signalling success, does not imply that the resulting cluster is without error. In this paper we explore the effect of errors in the fusion operation on microcluster construction. We will look at two different error models: (1) a non-equiprobable Pauli error model in which the surviving photon of a successful fusion operation undergoes a $\sigma_x$-rotation with probability $p_x$, a $\sigma_y$-rotation with probability $p_y$, a $\sigma_z$-rotation with probability $p_z$, where $\sigma_j$ is a Pauli spin operator. (2) An error in which a successful fusion operation is signalled though there is a probability that an incorrect projection was implemented by the PBS-waveplate combination. The PBS in an ideal successful fusion operation projects the incoming photons into states with the same polarization and thus one photon will exit each of the two output paths. However, an imperfect PBS may not project the photons into states with the same polarization but will nevertheless send the two photons out on different paths. In this way a successful fusion operation will be signalled despite the incorrect projection. Furthermore, the photons would actually be distinguishable based on polarization and thus the state of the unmeasured photons will depend on which of the two photons entering the fusion operation is actually measured. We paramaterize this error by $\alpha$ and the imperfect projection operator is then given as:
\begin{equation}
P_{PBS} = \left( \begin{array}{cccc}
1-\alpha & 0 & 0 & 0 \\
0 & \alpha & 0 & 0 \\ 
0 & 0 & \alpha & 0 \\
0 & 0 & 0 & 1-\alpha \\ \end{array} \right).
\end{equation}
The effects of both of these errors will be quantified via the fidelity of the constructed microclusters. 

We then fuse together imperfect microclusters and determine the accuracy of constructing the fused microclusters also via a fidelity. We note that other aspects of realistic photonic cluster state construction including the need to store cluster states (though imperfectly) during construction and effects of dephasing have been explored elsewhere \cite{YSW1,YSW2,WKK}.
 
\section{Microcluster Construction}

Our first step is the construction of the microcluster in each of the error environments. We assume an inexhaustible supply of perfect EPR pairs and the construction follows Fig. 1. In the non-equiprobable error environment we notice that $\sigma_x$ and $\sigma_y$ errors are symmetric. Thus, we set $p = p_x = p_y$. The fidelity of the constructed microclusters in this environment is shown in Table~\ref{tab:microPPzFide} as a function of the number of microcluster leaves. 

\begin{table}
  \begin{center}
    \begin{tabular}{| c | c |}
    \hline
    Leaves & Fidelity  \\ \hline
    1 & $1$  \\ \hline
    2 & $- q - pz$  \\ \hline
    3 & $q^2 + 2q p_z + 2 p_z^2$  \\ \hline
    4 & $-q^3 - 3q^2 p_z - 6q p_z^2 - 4 p_z^3$  \\ \hline
    5 & $q^4 + 4q^3 p_z + 12 q^2 p_z^2 + 16q p_z^3 + 8 p_z^4$  \\ \hline
    6 & $-q^5 - 5q^4 p_z - 20q^3 p_z^2 - 40q^2 p_z^3 + 40q p_z^4 - 16 p_z^5$  \\ \hline
    \end{tabular}
  \end{center}
\caption{Fidelity of microclusters constructed in the non-equiprobable error environment as a function of the number of leaves, with $p = p_x = p_y$ and $q = 2p-1$.}
  \label{tab:microPPzFide}
\end{table}

Interestingly, when the fidelity is factored into powers of $q =2p-1$ and $p_z$, the coefficients fall into a precise pattern based on the square array of transforms of binomial coefficients. The $(n,k)$ entry of the square array of transforms of binomial coefficients is given by $2^{n-1}(n+k)!/(n! \times k!)$. The coefficients of the fidelities are the antidiagonals of this table. Some entries are shown in Table~\ref{tab:transforms}.
\begin{table}
  \begin{center}
    \begin{tabular}{c c c c c}
    1 & 1 & 2 & 4 & 8 \\
    1 & 2 & 6 & 16 & 40 \\
    1 & 3 & 12 & 40 & 120 \\
    1 & 4 & 20 & 80 & 280 \\
    1 & 5 & 30 & 140 & 560 \\
    \end{tabular}
  \end{center}
  \caption{The first five rows and colums of the square array of transforms of binomial coefficients. Note that the antidiagonals of this table give the coefficients of the microcluster fidelity when built in the non-equiprobable error environment shown in Table~\ref{tab:microPPzFide}.}
  \label{tab:transforms}
\end{table}
This result suggests that the fidelity of a microcluster with an arbitrary number of leaves constructed in the non-equiprobable Pauli error environment can be read off of the table. 

In the equiprobable Pauli error environment, $p_x = p_y = p_z = p$, we find that the fidelity to first order in error probability is given by $1-3(n-1)p$ where $n$ is the number of microcluster leaves. 

An imperfect PBS may lead to the following error. One photon emerges from each of the exit paths of the PBS, thus a successful fusion operation will be signalled, but the PBS did not project the photons into the same polarization state. When building microclusters under these conditions a two-leaf microcluster has a fidelity given by:
\begin{equation}
F_2(\alpha) = \frac{(1-\alpha)^2}{1+2(\alpha^2-\alpha)}.
\end{equation} 
Additional leaves reduce the fidelity by factors of $F_2(\alpha)$ such that
\begin{equation}
F_n(\alpha) = F_2(\alpha)^{n-1}.
\end{equation}

Another possible fusion operation error is an imperfect rotation of the photon that exits the PBS towards the photon detector. This causes an undesirable (unitary) rotation of the other qubit taking part in the fusion. If the error is known, for example via experimental tomography of the fusion operation, it can be perfectly corrected. If it is unknown typical error correction techniques will have to be applied.   

\section{Microcluster fusion}

The next step is to fuse two microclusters together such that the two root nodes become nearest neighbors. This is done by attempting fusion with pairs of leaf nodes as shown in Fig.~\ref{micro}. Upon success, the root nodes are separated from each other by one leaf node (now attached to both roots). The root nodes can be made nearest neighbors by measuring the leaf node along the $y$-axis. We would like to know the fidelity of the two root nodes when the entire process is performed in the identified error environments as a function of the number of leaves initially on the microclusters. In addition, because the errors and fusion operations do not commute, which of the fusion operations between leaves is successful will also effect the fidelity of the resulting two qubit cluster. In a real system the microclusters would likely be stored in some imperfect storage unit in between fusion attempts lowering the cluster state fidelity as explained in \cite{YSW1}. Here we assume no errors except those associated with the fusion operations. We also assume a perfect $y$-axis measurement. To isolate the root nodes we measure all extraneous leaves along the $z$-axis. 

\begin{figure}
\begin{center}
\includegraphics[height=4cm]{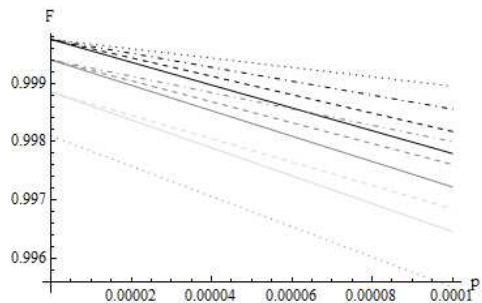}
\caption{\label{fusedfids}
Fidelity of two root qubit clusters as a function of $p = p_x = p_y = p_z$ for $\alpha = .01$. Plots are shown for when the first fusion attempt is successful (black), the second (gray), the third (light gray) and the fourth (lowest). The number of initial leaves on the microcluster is 2 (black, dotted), 3 (chain), 4 (dashed), and 5 (solid). }
\end{center}
\end{figure}

In Table~\ref{tab:microFusePzEFideComp} and Figure~\ref{fusedfids} we show the fidelity of the two root nodes of the fused microslusters as a function of the number of initial leaves on the cluster and which leaf fusion operation was successful. We assume that the first fusion attempt is done with the newest leaf and subsequent attempts use gradually older leaves. Extraneous leaves are removed by measurement along the $z$-axis.

\begin{table}
\begin{center}
\begin{tabular}{|c|c|c|c|c|}
\hline
Leaves & 1 & 2 & 3 & 4  \\ \hline
1 & $1-\frac{5\alpha^2}{2}-8p$ & failure & failure & failure \\ 
  & $+28\alpha^2p$ & & & \\\hline
2 & $1-\frac{5\alpha^2}{2}-12p$ & $1-6\alpha^2-14p$ & failure & failure \\ 
  & $+8\alpha p+40\alpha^2p$ & $+142\alpha^2p$ & & \\\hline
3 & $1-\frac{5\alpha^2}{2}-16p$ & $1-6\alpha^2-18p$ & $1-\frac{23\alpha^2}{2}-20p$& failure \\ 
  & $+24\alpha p+28\alpha^2p$ & $+8\alpha p+192\alpha^2p$ & $+424\alpha^2p$ & \\\hline
4 & $1-\frac{5\alpha^2}{2}-20p$ & $1-6\alpha^2-22p$ & $1-\frac{23\alpha^2}{2}-24p$ & $1-19\alpha^2-26p$ \\ 
  & $+48\alpha p-24\alpha^2p$ & $+24\alpha p+218\alpha^2p$ & $+8\alpha p +536\alpha^2p$ & $+954\alpha^2p$ \\\hline
\end{tabular}
\end{center}
\caption{Fidelities for 1 through 4 leaves with $p_x = p_y = p_z = p$ for two root nodes of fused microclusters depending on which  fusion operation 1 through 4 is successful. Only terms to first order in $p$ and second order in $\alpha$ are shown. Failure refers to when there are insufficient to achieve a successful fusion.}
\label{tab:microFusePzEFideComp}
\end{table}

We see from the fidelity results that for $\alpha/p \gg 1$ the factor most effecting the fidelity is which fusion operation is successful. The initial number of leaves on the microcluster is unimportant. As $\alpha/p$ gets smaller the number of leaves plays a more prominent role in the fidelity. The more leaves initially on the cluster the lower the fidelity.  

\section{Conclusions}

In this work we have explored the construction and fusion of microclusters, basic building blocks for fault tolerant cluster state quantum computation, in two different noisy environments. The first assumes that after successful fusion, the unmeasured qubit is subject to a nonequiprobable Pauli error. The second assumes an imperfect polarizing beam splitter. In both cases adding more leaves to the microcluster lowers the mircocluster fidelity. 

The reason to use microclusters is to ensure fusion probability approaching one by fusing leaf nodes rather than information bearing root nodes. Thus, we fuse together two microclusters and determine the resulting fidelity of the two root nodes. This fidelity is also a function of the number of leaves on the intial microclusters and which fusion operation between leaves is successful. These simulation are important for purposes of fault tolerant cluster state quantum computation.       

\section{Acknowledgements}
It is a pleasure to acknowledge useful conversations with S. Pappas and support from the MITRE Innovation Program. 


\end{document}